# PoW, PoS, & Hybrid protocols: A Matter of Complexity?

Renato P. dos Santos

Blockchain Researcher at *ULBRA – The Lutheran University of Brazil*
renatopsantos@ulbra.edu.br.

Melanie Swan

Technology Theorist and Founder at Institute for Blockchain Studies
melanie@BlockchainStudies.org

**Abstract**

In a previous paper, it was discussed whether Bitcoin and/or its blockchain could be considered a complex system and, if so, whether a chaotic one, a positive response raising concerns about the likelihood of Bitcoin/blockchain entering a chaotic regime, with catastrophic consequences for financial systems based on it. This paper intends to simplify and extend that analysis to other PoW, PoS, and hybrid protocol-based cryptocurrencies. As before, this study was carried out with the help of Information Theory of Complex Systems, in general, and Crutchfield's Statistical Complexity measure, in particular. This paper is a work-in-progress. Whereas PoW consensus was shown to be highly non-complex, the Nxt PoS consensus method studied shows an outstandingly higher measure of complexity, which is undesirable because it introduces unnecessary complexity into what should be a simple computational system. This paper is a work-in-progress and undoubtedly prone to incorrectness as a few cryptocurrencies may have changed their consensus algorithms. As next step, we intend to uncover some other measures that capture the qualitative notion of complexity of systems that can be applied to these cryptocurrencies to compare with the results here obtained. As a final thought, however, considering that a certain amount of chaoticity may have been potentially introduced in the Bitcoin market by the presence of the capital gains-seekers, one could wonder whether the recent surge of blockchain technology-based start-ups, even discounting all the scam cases, could not help to reduce non-linearity and prevent chaos.

## Introduction

We are undoubtedly witnessing a surge of cryptocurrencies that use cryptographic protocols and algorithms to secure their transactions and the creation of additional units. Bitcoin, created under the mysterious Japanese pseudonym of Satoshi Nakamoto (2008), besides being the first one, still is by far the best-known one[1].

---

[1] The widespread practice will be followed here of distinguishing between "Bitcoin" (singular with an uppercase letter *B*), labelling the protocol, software, and community, and "bitcoins" (with a lowercase *b*), labelling units of the currency, which is represented as BTC or a capital letter *B* with two vertical lines going through it, a symbol created by Satoshi Nakamoto.

While the concept of decentralised digital currency, as well as alternative applications like property registries, has been around for decades, Satoshi Nakamoto's consensus algorithm, known as "proof of work" (PoW), was a breakthrough because it simultaneously provided:
1. A simple and moderately effective consensus algorithm for collective agreement on a set of updates to the state of the Bitcoin ledger.
2. A mechanism that both allowed free entry into the consensus process and prevented Sybil attacks. (Buterin, 2014).

Nakamoto's consensus algorithm does this by substituting a formal barrier to participation by an economic one – the weight of a single node in the consensus voting process is directly proportional to the computing power that the node brings (Buterin, 2014).

Before proceeding with our analysis, it must be understood that the blockchain is the result of "the asynchronous interaction of a resilient network of thousands of uncomplicated, independent nodes, all following straightforward, algorithmic rules to accomplish a myriad of financial processes" (Antonopoulos, 2014, p. 177). Notwithstanding, some authors assume that the whole blockchain code is characterised by a high degree of complexity, apparently confusing the high complication (of the code) with an eventual complexity of the resulting blockchain and Bitcoin ecosystem. Consequently, blockchain seems to worth an analysis through Complexity Theory to clarify this issue.

To obscure things further, there is no concise definition of a complex system, which has been oft identified with complicated or random (stochastic) systems. Besides, the term "complexity" has been so much used by so many authors, both scientific and non-scientific, that the word has, unfortunately, almost lost its meaning (Feldman & Crutchfield, 1998).

Another remark that should be made is on attempts of understanding chaoticity in terms of high volatility. Volatility is usually defined as a statistical measure of dispersion **around the average** of any random variable such as market parameters (Belev & Todorov, 2015), assuming that its price follows a Gaussian random walk or a similar distribution and that some sort of regression toward the mean always happens. It should be noticed, however, that Mandelbrot showed long ago that financial markets are characterised by "wild randomness," in the sense that the price changes do not follow a Gaussian distribution, but rather Lévy stable distributions **having infinite variance** (Mandelbrot, 1963). Therefore, if some market enters a chaotic regime, there will be nothing predictable about it, and the prices can go anywhere. In other words, ordinary volatility is expected and even desirable to some degree as it provides profitable opportunities; **chaoticity is a menace to any market**.

Consequently, as done in the previous paper (dos Santos, 2017), the systems' complexities is analysed here by means of the Crutchfield's Statistical measure of complexity (Feldman & Crutchfield, 1998). This choice is justified as Ladyman et al. (2013) showed that, among the various measures of complexity available in the scientific literature, it was one that best captures the qualitative notion of complexity of systems.

From a technical standpoint (Wood, 2014), any PoW-based cryptocurrency such as Bitcoin or Ethereum can be viewed as a *state-transition system*, where there is a "state" consisting of the ownership status of all existing tokens and a "state-transition function" that takes the previous state $\sigma_{t-1}$ and a valid transaction $T$ and outputs a new state $\sigma_t$ as the result. Formally,

$$\sigma_t \equiv Y(\sigma_{t-1}, T)$$

where $Y$ is the state-transition function.

More specifically, as shown in the previous paper (dos Santos, 2017), Bitcoin blockchain can be seen as an *infinite-string-production ϵ-machine* (Figure 1) that oscillates between two states about every 10 minutes:

1. $\sigma_m$ (mining state): A new block was just incorporated into the blockchain, and the machine starts mining a new block that includes most of the pending transactions collected from around the world into the *transaction pool*. Hashes are generated and tested against the network's *difficulty target*.

2. $\sigma_b$ (broadcasting state): A nonce that results in a hash smaller than the target is found, the validated block is broadcast to the P2P network for inclusion into the blockchain. If a blockchain *fork* (Antonopoulos, 2014, p. 199) happens, the global Bitcoin network ultimately converges to a new consistent mining state.

*Figure 1*. Bitcoin blockchain as seen as an infinite-string-production ϵ-machine that oscillates between two states $\sigma_m$ and $\sigma_b$. The inscribed circle indicates the start-state corresponding to the generation of the origin block.

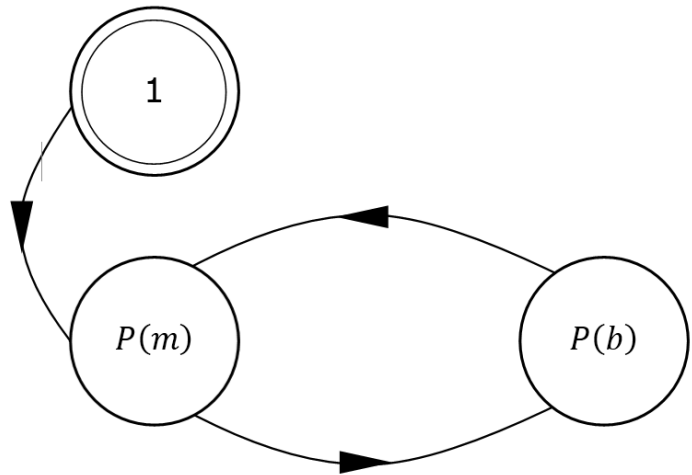

The Crutchfield's Statistical Complexity measure (Crutchfield, 2012) of Bitcoin's blockchain was previously estimated (dos Santos, 2017), using the usual "number of leading zero bytes in the hash" approximation, for the then current network production rate of 4,265,775.24 Tera ($10^{12}$) hashes per second, as $C_\mu \cong 1.56 \times 10^{-20}$.

A more straightforward and yet more accurate procedure is to consider that new blocks are created about every 10 minutes (Antonopoulos, 2014, p. 2), that is 600 seconds, and, consequently, for the same network production rate value, it takes in average $(4.27 \times 10^{18}) \times 600 \cong 2.56 \times 10^{21}$ hashes to find one that is lower than the target[2]. Therefore, the probability of the broadcasting state is $P(\sigma_b) = 1/(2.56 \times 10^{21}) \cong 3.9 \times 10^{-22}$, while that of the mining state $\sigma_m$ is $P(\sigma_m) = 1 - P(\sigma_b) \cong 1 - 3.9 \times 10^{-22}$, the statistical complexity $C_\mu$ resulting[3]

$$C_\mu = -\left((1 - 3.9 \times 10^{-22})\log_2(1 - 3.9 \times 10^{-22}) + (3.9 \times 10^{-22})\log_2(3.9 \times 10^{-22})\right)$$

$$C_\mu \cong 2.83 \times 10^{-20},$$

a value that is almost twice the one obtained before but is of the same order of it and leads to the very same conclusion that Bitcoin blockchain may hardly be considered a complex system.

---
[2] The target is calculated by dividing the maximum target used by SHA-256 (which should logically correspond to 256 1 binary digits, but, because Bitcoin stores the target as a floating-point type, this is truncated to approximately $2^{224}$, which can be represented as the 64-hex hash 0x00000000FFFF0000000000000000000000000000000000000000000000000000, being "0x" the usual prefix to flag hexadecimal numerals) by the difficulty.

Straightforwardly applying this procedure to a few others PoW cryptocurrencies, we obtain the results shown in Table 1.

Table 1. *Crutchfield's Statistical Complexity measure $C_\mu$ calculated for a few PoW cryptocurrencies.*

| Currency | Bitcoin | Ether | Bit. Cash | BTC Gold | Litecoin | Dash | Monero | Eth. Classic |
|---|---|---|---|---|---|---|---|---|
| Block time† | 10 min | 0,25 min | 10 min | 10 min | 2,5 min | 2,5 min | 2 min | 0,25 min |
| Hashrate (hash/s) † | 2,78×10$^{19}$ | 2,77×10$^{+14}$ | 4,62×10$^{+18}$ | 3,50×10$^{+07}$ | 2,98×10$^{+14}$ | 1,80×10$^{+15}$ | 4,32×10$^{+08}$ | 7,70×10$^{+12}$ |
| $C_\mu$ | 4,51×10$^{-21}$ | 1,28×10$^{-14}$ | 2,62×10$^{-20}$ | 1,70×10$^{-09}$ | 1,27×10$^{-15}$ | 2,19×10$^{-16}$ | 7,15×10$^{-10}$ | 4,17×10$^{-13}$ |

| Currency | Zcash | Vertcoin | Dogecoin | Feathercoin | BlackCoin | Namecoin | Auroracoin |
|---|---|---|---|---|---|---|---|
| Block time† | 2,5 min | 2,5 min | 1 min | 1 min | 1 min | 10 min | 1 min |
| Hashrate (hash/s) † | 4,23×10$^{+08}$ | 1,02×10$^{+12}$ | 2,21×10$^{+14}$ | 5,45×10$^{+09}$ | 9,61×10$^{+13}$ | 2,20×10$^{+19}$ | 1,37×10$^{+15}$ |
| $C_\mu$ | 5,88×10$^{-10}$ | 3,18×10$^{-13}$ | 4,14×10$^{-15}$ | 1,21×10$^{-10}$ | 9,33×10$^{-15}$ | 5,68×10$^{-21}$ | 7,00×10$^{-16}$ |

**† Obtained on May 12, 2018, from [https://bitinfocharts.com](https://bitinfocharts.com)**

These extremely low statistical complexity results lead us to the conclusion that PoW-based blockchains, in general, can hardly be considered complex, confirming and extending to all these cryptocurrencies Nakamoto's statement about Bitcoin that "the network is robust in its unstructured simplicity" (2008). The functioning of these blockchains may be regarded as algorithmically complicated, but not complex.

PoW was undoubtedly crucial to give birth to Nakamoto's major breakthrough. However, its computing-power-intensive nature means that PoW crypto-currencies are dependent on energy consumption, which introduces a significant cost overhead in the operation of those networks that is borne by the users via a combination of inflation and transaction fees. As the mint rate slows in Bitcoin network, reducing the block reward that the successful miner takes, it is putting pressure on raising transaction fees to sustain a preferred level of security, as predicted by King & Nadal (2012).

---

[3] Due to $P(\sigma_b)$ being much smaller than 1, we used [Padé Approximation](#) $\log_e(1 - x) \cong -x(6 - x)/(6 - 4x)$ to increase the precision of calculation of the $(1 - P(\sigma_2))\log_2(1 - P(\sigma_2))$ term.

For this reason, there was recently a burst of popularity of the cryptocurrencies that used an alternative algorithm known as "Proof-of-Stake" (PoS) for choosing the block creators. Unlike the PoW-based cryptocurrencies, where miners must solve complicated cryptographical puzzles to be able to create blocks and be rewarded for it, in PoS-based cryptocurrencies the creator of the next block is chosen in a deterministic (pseudo-random) way, and the chance that an account is selected depends on its wealth (the stake). In other words, PoS calculates the weight of a node as being proportional to its currency holdings and not its computational resources (Buterin, 2014). For this reason, in PoS cryptocurrencies, the blocks are usually said to be forged (in the blacksmith sense of this word), or minted (Popov, 2016).

Pure PoS cryptocurrencies such as Nxt choose the account that has the right to generate the next block according to the number of coins in the account; the wealthier the account is, the higher is the chance that it will be able to generate the next block and receive the corresponding transaction fees. Habitually, one assumes that this probability should be precisely proportional to the account's balance, although it is not quite true for the Nxt (Popov, 2016).

In Nxt, to participate in the block forging process, each active account $k$ extracts the first 8 bytes of the result of applying 8 times in sequence a SHA256 hashing function to its account public key and the generating signature of the current block, a value that is referred to as this particular account *hit* $H_k$. As it depends on the account public key, it varies with the account attempting to forge on top of a specific block. Even if no pseudo-random number generators are involved here, as the result of a hash function is virtually unpredictable (NIST, 2012), it is, nevertheless, still reasonable to regard the hits $H_k$ as i.i.d. random variables[4] with uniform distribution on that interval (Nxt community, n.d.; Nxt Wiki contributors, 2017).

---

[4] Yu and Wang announced at the 'rump session' of Eurocrypt 2008 that they had shown non-randomness for SHA-256, but no details have been published to date.

Furthermore, each account $k$ calculates its own *target* value, based on its current effective balance as

$$T_k = T_b \times S \times B_k$$

where:

- $T_k$ is the new target value for the $k^{th}$ account. It grows with each second that passes since the timestamp of the previous block, limited by a check hard-coded within the Nxt protocol to a maximum value of $2^{64}/(2 \times 60) = 1.53722867310^{17}$ and a minimum of one half of the last block base target value (Nxt Wiki contributors, 2016).
- $T_b$ is the base target value, usually expressed as percentage of the base target of the genesis block (153722867.3), varies from block to block, and is derived from the previous block base target $T_p$, $\acute{S}$, the average of times that were required to generate the last 3 blocks, and three constants $maxratio$, $minratio$, and $\gamma$, using a formula that smoothly increases or reduces the new base target depending on the previous block having taken less or more than one minute to be generated, therefore ensuring a generation block time between blocks of 60 seconds in average (Nxt community, n.d.).
- $S$ is the time passed, in seconds, since the last block has been generated. It is the same for all accounts.
- $B_k$ is the effective balance (stake) of the $k^{th}$ account. To avoid shuffling attacks, only the amount of *at least* 1000 NXT that has been confirmed at least 1440 times on the last 24 hours counts towards this balance (Nxt community, n.d.; Nxt Wiki contributors, 2017).

To merit the right to forge (generate) a block, each active Nxt account $k$ "compete" by waiting the $S$ factor in the target formula above being increased with each passing second that no block is generated, until its particular target value $T_k$ surpasses its own random hit value $H_k$. Notice that the bigger the account stake $B_k$ is, the higher and fast-growing will be its target $T_k$, making it easier to surpass its hit $H_k$ (Nxt community, n.d.; Nxt Wiki contributors, 2017). In other words, in Nxt, an account's "chance" to forge a block hinges only on its current 'stake' (which is a property of each account), the time since the last block (which is shared by all forging accounts) and the base target value (which is also shared by all accounts) (Nxt community, n.d.).

Differently from Bitcoin, then, instead of a global target against which nodes keep mining their nonces until one is found that is less than the target, in Nxt, the individual hits are calculated beforehand, and new target increased values are generated each second until one satisfying the $hit < target$ condition is found. (Andruiman, 2014).

Therefore, PoS-based blockchains can also be seen as an infinite-string-production machine that oscillates between two states about every minute:

1. $\sigma_t$ (targeting state): A new block was just incorporated into the blockchain. Each active forging account $k$ generates its own random hit value $H_k$ and starts generating new, increasing individual target value $T_k$ each second that no block is generated, until some of them surpass their own hit values.
2. $\sigma_b$ (broadcasting state): A few accounts hit their own target and win the right to forge candidate blocks. Each one of them bundles up to 255 unconfirmed transactions into a new block along with all its required parameters and broadcasts it to the network as a candidate for the blockchain. If multiple candidate blocks were generated, the block with the highest *cumulative difficulty* value will ultimately win and be inserted on the top of the blockchain.

Consequently, applying to Nxt the same procedure to evaluate its statistical complexity $C_\mu$, as a block is forged about every 60 seconds, the probability of a node calculating an individual target that is bigger than its hit is $P(\sigma_t) = 1/60 \cong 1.67 \times 10^{-2}$ and $C_\mu \cong 0.122$, a value of Crutchfield's Statistical Complexity measure that is 10 to 20 orders of magnitude bigger than those of Table 1 and that, consequently, raise serious concerns about the possibility of Nxt entering a chaotic regime at any time without notice.

There are also other protocols PoS with conceptually different implementations. For example, the forging probability may also depend on the time the coins were in the account without being transferred (the so-called *coin age*) (Popov, 2016). Coin age can be simply defined as currency amount times holding period. In a simple to understand example provided by King and Nadal (2012), if Bob receives 10 coins from Alice and holds it for 90 days, one can say that Bob accumulates 900 coin-days of coin age. Additionally, when Bob spends the 10 coins he received from Alice, one says that the coin age Bob accumulated with these 10 coins had been *consumed* (or *destroyed*).

The concept of coin age actually was known to Nakamoto at least as early as 2010 and used in Bitcoin to help prioritise transactions, for example, although it did not play much of a critical role in present Bitcoin's security model. Scott Nadal and Sunny King (a pseudonym) independently rediscovered the concepts of PoS and coin age in October 2011, whereby realising that PoS could indeed replace most PoW's functions with a careful redesign of Bitcoin's minting and security model (King & Nadal, 2012).

Another example of PoS variation is Reddcoin's Proof of Stake Velocity (PoSV), which intends "to encourage both ownership (Stake) and activity (Velocity) which directly correspond to the two main functions of Reddcoin as a real currency: a store of value and a medium of exchange." (Ren, 2014)

There is also hybrid PoW+PoS implementations, in which PoW mining works as both a steady distribution channel for the cryptocurrency and a fall-back network security mechanism. As PoW block rewards go down over time, the PoS protocol has enough time to move to the spotlight (Ren, 2014).

For example, in King and Nadal's Peercoin design, a new minting process is introduced for PoS blocks in addition to Bitcoin's PoW minting, and blocks are separated into two distinct types: PoW blocks and PoS blocks. The PoS in the new type of blocks is a special transaction called *coinstake* (named after Bitcoin's special transaction *coinbase*). In the coinstake transaction, block owner pays himself thereby consuming his coin age, while gaining the privilege of generating a block for the network and minting for PoS. The first input of coinstake is called *kernel* and is required to meet a specific hash target protocol, thus making the generation of PoS blocks a stochastic process similar to PoW blocks. However, a significant difference is that the hashing operation is done over a limited search space (more specifically one hash per unspent wallet-output per second) instead of an unlimited search space as in PoW. Thus no significant consumption of energy is involved (King & Nadal, 2012).

The hash target that stake kernel must meet is a target per unit coin age (coin-day) consumed in the kernel (in contrast to Bitcoin's PoW target which is a fixed target value applying to every node). Thus, the more coin age consumed in the kernel, the easier meeting the hash target protocol. For example, if Bob has a wallet-output which accumulated 100 coin-years and expects it to generate a kernel in 2 days, then Alice can roughly expect her 200 coin-year wallet-output to generate a kernel in 1 day. In Peercoin design, both PoW hash target and PoS hash target are adjusted continuously rather than Bitcoin's two-week adjustment interval, to avoid a sudden jump in network generation rate (King & Nadal, 2012).

Table 2 exhibits the results of applying this procedure to a few PoS or hybrid cryptocurrencies.

Table 2. *Crutchfield's Statistical Complexity measure $C_\mu$ calculated for a few PoS or hybrid cryptocurrencies.*

| Currency | NXT | Reddcoin | Peercoin | BlackCoin | NovaCoin |
|---|---|---|---|---|---|
| **Block time†** | 60 s | 1 min | 10 min | 60 s | 10 min |
| **Hashrate (hash/s) †** | 1,0 | $1,30 \times 10^{+10}$ | $3,32 \times 10^{+16}$ | $1,06 \times 10^{+14}$ | $4,42 \times 10^{+11}$ |
| $C_\mu$ | 0,122 | $5,26 \times 10^{-11}$ | $3,29 \times 10^{-18}$ | $8,52 \times 10^{-15}$ | $1,86 \times 10^{-13}$ |

† Obtained on May 12, 2018, from https://bitinfocharts.com

It is noticeable how the high difficulty and consequently needed higher hashrate of the PoW part of the protocol contributes to the lower value of the complexity measure in comparison to the Nxt PoS protocol, for the reasons discussed above. As the time interval between blocks forging in Nxt is kept around 60 seconds, the probability of any node to be selected is hugely higher than other currencies. This may be not a feature of PoS protocol itself, however, but a characteristic of Nxt. It is conceivable, therefore, that some different implementation of the basic PoS protocol could have a higher competition among 'forgers' and, consequently, a lower complexity.

**Conclusion**

This work suggests that Crutchfield's Statistical Complexity may be used as an effective analysis tool to evaluate the viability of proposed high-performance network cryptographic methods from the available quantitative data.

Whereas PoW consensus was generally shown to be highly non-complex, Nxt PoS consensus protocol shows an outstandingly higher measure of complexity. This feature is undesirable because it introduces unnecessary complexity into what should be a simple computational system, and therefore, the proposed PoS methods seem to be more convoluted and complicated to be globally scalable, applicable, and sustainable as a model of decentralised network computing. This high complexity does not seem to come from the PoS concept per se but from the smaller competition among forgers, derived from the tremendously smaller number of trials per second are made in Nxt to select the next forger.

This paper is a work-in-progress and undoubtedly prone to incorrections as a few cryptocurrencies may have changed their consensus algorithms. As next steps, we intend to extend this analysis to other PoS-based cryptocurrencies to investigate whether this higher complexity generalises to PoS systems, as well to apply this research to next-gen consensus algorithms such as DFINITY, MimbleWimble, DAG, HashGraph, and IOTA. Furthermore, we will try to uncover a more extensive suite of "complexity math" approaches for the assessment of the complexity of cryptocurrencies to validate the results here obtained.

As a final thought, returning to the discussion at the end of previous paper (dos Santos, 2017), in which Siddiqi (2014) attributes a potential chaoticity in the Bitcoin market to the non-linearity introduced in it by the presence of the capital gains-seekers, one could wonder whether the recent surge of blockchain technology-based start-ups, even discounting all the ICO scam cases, could not help to reduce non-linearity and prevent chaos.